\documentclass[twoclomn,physicab]{elsarticle}%

\usepackage{amsmath}
\usepackage{amsfonts}
\usepackage{amssymb}
\usepackage{graphicx}%
\providecommand{\U}[1]{\protect\rule{.1in}{.1in}}

\journal{Physica B}
\begin{document}
\begin{frontmatter}
\title{Magnetic field noise analyses generated by the interactions between a nitrogen
vacancy center diamond and surface and bulk impurities}
\author{Philip Chrostoski$^{\ast,\dag,\ddag}$}
\author{Bruce Barrios$^{\ast}$}
\author{D. H. Santamore$^{\ast}$}
\address{$^{\ast}$Department of Physics and Engineering Physics, Delaware State
University, Dover, DE 19901, USA}
\address{$^{\dag}$Department of Physics and Astronomy, University of Missouri St.
Louis, St. Louis, MO 63121, USA}
\address{$^{\ddag}$Department of Physics, Missouri University of Science and
Technology, Rolla, MO 65409, USA}
\begin{abstract}
We investigated the mechanism of magnetic noise due to both surface and bulk impurities. For surface noise, we apply the Langevin method to spin fluctuation theory to calculate the noise for paramagnetic surface impurities absorbed in a thin layer of water. We find that the mechanisms generating noise are spin flip and spin precession which depend on impurity spin relaxation and spin precession time. For the bulk noise, we consider carbon-13 and nitrogen as impurities and employ the correlated-cluster expansion to calculate noise. Carbon-13 noise is a few orders of magnitude larger than nitrogen due to the higher impurity density in the typical NV center diamond system. We also find that the noise in the secular approximation underestimates noise under low applied magnetic field. Overall, the major source of magnetic field noise is spin precession noise, which is more than five orders of magnitude larger than the spin flip noise.
\end{abstract}
\end{frontmatter}


\section{Introduction}

Nitrogen vacancy (NV) center diamonds have recently emerged as promising
candidates for nanoscale sensors, because of the long quantum coherence times
of their spin states well above room temperature \cite{BNT2009} and because of
their high sensitivity to external magnetic and electric fields
\cite{hanson2016nature, wrachtrup2011nature} both at room temperature
\cite{acosta2010rev, toyli2013sci, beausoleil2013letters} and at very low
temperatures \cite{wrachtrup2015physical}. These features have led to the use
of NV centers as magnetic and electric field sensors \cite{hanson2016nature,
wrachtrup2011nature, degen2014chem}, quantum information \cite{mazeproperties}%
, magnetic imaging \cite{walsworth2013nature, hollenberg2017commun}, high
spatial resolution devices \cite{yacoby2017applied} biomarkers
\cite{chang2013nanotech}, and temperature sensors \cite{mikhail2013nature,
manson2014nano}. As a result, NV centers are a subject of intensive research,
not only in physics, but also in chemistry \cite{walsworth2015nanotechnol,
wrachtrup2013scie, lukin2017science}, life science
\cite{connolly2015natmethods}, and planetary studies \cite{kuan2014science}.

One of the current serious problems is noise, which reduces sensitivity of NV
center diamond sensors by broadening of the spectral linewidth and reduces
spectral resolution of the device \cite{kehayias2014microwave,
balasubramanian2014nitrogen}. Noise in NV center diamonds has two sources, the
electric and magnetic fields. In some crystals used for optical device
applications, crystal field splitting can produce Fano noise from charge
fluctuations that contributes to the electric field noise
\cite{Li17,Fan20,Raza20}. In previous work we studied the electric field noise
and found that surface electric field noise comes from surface charge
fluctuations where the surface covering layer is the key to controlling this
type of noise \cite{chrostoski2018electric}.

In this paper, we study magnetic field noise in an NV center diamond. This
noise arises from interactions of the NV center electron spin with impurity
spins in the system. The relevant impurities can occur both on the surface
\cite{OSV09, BVW72, THR13} and in the bulk \cite{knowles2014observing}. We
treat the surface impurity interactions in Sec.\ \ref{Sec_Surface}, and the
bulk impurity interactions in Sec.\ \ref{Sec_Bulk}.

The surface impurities we consider are absorbed molecules in a condensed thin
layer of water that can hop around randomly among the terminated surface
sites. We investigate hydrogen (H-), oxygen (O-), and fluorine (F-) terminated
systems, since these are commonly used in NV center experiments where the
noise reduction effects have been observed experimentally \ \cite{RDO2014,
LHG2015, HAU11, CHU13}, but the mechanisms and the key controlling factors of
noise generation have never been investigated theoretically.

There are two mechanisms for NV center electron spin fluctuations that can
contribute to the magnetic field noise: (a) fluctuations associated with the
spin precession and (b) fluctuations due to spin flips. We model the noise
using the Langevin method applied to the Bloch equation \cite{GI2012, MMG2015,
SGHGAB16, LAX66}. We obtain the noise spectra by using the Langevin approach
to spin fluctuation theory for all three different terminations.

We then study two typical bulk impurities in NV center diamonds: (a)
carbon-$13$ isotopes that occur naturally and constitute $1$\textrm{\%} of the
carbon atoms in the diamond; and (b) nitrogen atoms that remain after ion
implantation during the production of the NV centers \cite{laraoui2013high,
rabeau2006nitrogen}. Both species have nuclear spins that interact with the NV
center electron spin, leading to magnetic field fluctuation noise
\cite{pezzagna2010creation, mrozek2015longitudinal, takahashi2008quenching}.
We calculate the noise spectra using the correlated-cluster expansion method
\cite{yang2008quantum}, which is particularly suited for dense systems of
interacting particles \cite{yang2008quantum, zhao2012decoherence,
yang2009quantum}. We also examine the applicability of the secular
approximation by comparing it with non-secular exact calculations.

In Sec.\ \ref{subsec_SurfaceNoiseDiscussion}, we discuss the results of the
surface impurity noise calculations. The O-terminated surface gives a much
smaller noise floor than the other two terminated surfaces at the same amount
of impurities. This is because of the very short (picosecond) spin-lattice
relaxation time \cite{THK01} of the impurity electron spins in the
O-terminated system, compared to the H- and F-terminated systems, whose
spin-lattice relaxation times are on the order of microseconds \cite{SOI94,
RAG81}.

A short relaxation time leads to a short impurity-impurity spin correlation
time resulting in spin flip fluctuation noise only. On the other hand, a long
relaxation time leads to a long impurity-impurity spin correlation time, which
results in activating another effect of spin-precession fluctuation noise to
the already existing spin flip fluctuation noise. The spin precession
fluctuation noise is larger than the spin flip fluctuation noise. The bonding
between the hopping terminating atoms and the dangling bond will be reaction
limited. This allows most of the sites above the NV center to be occupied by a
terminating atom. At this point, the terminating atoms with long relaxation
times are still generating spin precession noise. At high frequencies (short
time scales), the spin precession is less prevalent as not enough time has
passed for them to bond with dangling bonds.

The effective relaxation time is limited by whichever is faster of the
spin-lattice relaxation rate or the \textquotedblleft
hopping\textquotedblright\ (i.e., the impurities moving around the surface)
rate, yet the number of impurities affects only the hopping rate. Thus the
relaxation time of H- and F-terminated surfaces transitions from being
controlled by the hopping rate to being controlled by the spin flip rate as
the impurity number increases. In contrast, the relaxation time of
O-terminated surfaces is controlled by the spin flip rate for all practical
levels of impurity incidence.

In Sec.\ \ref{subsec_BulkNoiseDiscussion}, we discuss the results of the bulk
impurity noise calculations. We find that the carbon-$13$ nuclear spin
interaction is roughly two orders of magnitude larger than that of nitrogen.
This is mainly because the number density of carbon in NV center diamonds
exceeds that of nitrogen. As a result, the spacing between the NV center and
the carbon-$13$ atoms will be smaller than the spacing between the NV center
and the nitrogen atoms, resulting in a stronger interaction between the NV
center and the carbon. In addition, the total fluctuations caused by the
carbon-$13$ nuclear spin flip-flops are larger than those caused by the
nitrogen nuclear spins.

The secular approximation, which is related to the rotating wave
approximation, has been commonly used for the calculations of bulk impurity
interactions as it is simpler and easy to evaluate. However, under low applied
magnetic field ($B<1$ $\mathrm{G}$), we find that the secular approximation
underestimates the amount of noise by a factor of $2.7$. The secular
approximation ignores the off-diagonal terms of the nuclear dipole-dipole
interaction matrix, which is a valid assumption for the higher applied
magnetic field ($B>100$ $\mathrm{G}$). Those matrix elements are are not
negligible for low magnetic field: in fact, they account for our factor of
$2.7$ discrepancy between the secular and non-secular calculations. Thus, the
applied magnetic field affects the amount of noise: the higher the applied
magnetic field, the less noise as the off-diagonal terms become insignificant.
This theoretical result agrees with experimental reports \cite{BSS17, HKS2016}.

\section{Surface impurity interactions\label{Sec_Surface}}

\subsection{Model\label{subsec_surfacemodel}}

\begin{figure}[ptb]
\begin{center}
\includegraphics[width=10 cm] {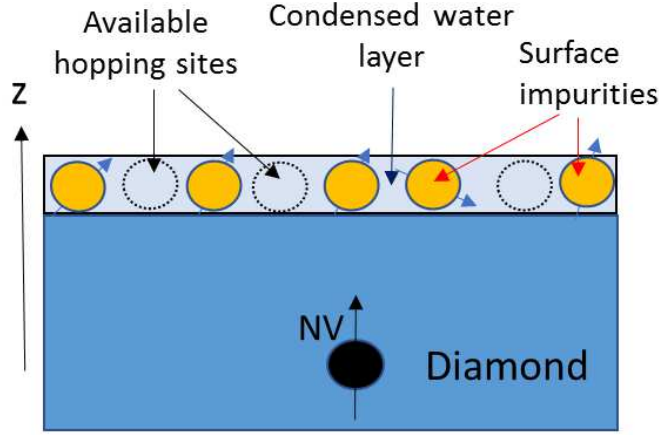}
\end{center}
\caption{Our surface impurity model. Paramagnetic surface impurities are
absorbed on the condensed water monolayer of hydrogen (H-), oxygen (O-), and
fluorine (F-) terminated surfaces. These impurities can hop around among the
terminated surface sites.}%
\label{fig1_Surfacemodel}%
\end{figure}

We first calculate the magnetic field noise from the surface impurity
interactions. The magnetic field noise from the surface impurities has two
sources: (a) fluctuations associated with an energy transfer at the spin
precession frequency and (b) fluctuations due to spin flip induced when the
impurity hops to a new termination site. Fluctuations of both types affect the
dipole-dipole interactions between the impurities and the NV centers, and lead
to noise.

As shown in Fig.\ \ref{fig1_Surfacemodel}, we model the paramagnetic surface
impurities in the condensed thin layer of water on hydrogen (H-), oxygen (O-),
and fluorine (F-) terminated systems. These impurities can move around
randomly (``hop") among the termination sites.

The Hamiltonian of the system is \cite{SGHGAB16}:%
\begin{equation}
H=H_{0}+\hbar\left(  \mathbf{\hat{\Omega}}_{B}+\mathbf{\hat{\Omega}}%
_{imp}\right)  \cdot\mathbf{S}. \label{Hamiltonian_surf}%
\end{equation}
Here $H_{0}$ is the free Hamiltonian for the NV center and surface impurity
electrons. The NV electron spin precession due to the applied external
magnetic field is expressed in%
\begin{equation}
\mathbf{\Omega}_{B}=\mu_{B}g_{e\left(  \parallel,\perp\right)  }%
\mathbf{B}=\frac{\mu_{B}}{\hbar}\left[  g_{e}^{\parallel}B_{z}\mathbf{\hat{z}%
}+g_{e}^{\perp}\left(  B_{x}\mathbf{\hat{x}}+B_{y}\mathbf{\hat{y}}\right)
\right]  , \label{Omega_B}%
\end{equation}
in which $\mathbf{B}\left(  x,y,z\right)  $ is the external magnetic field and
$g_{e}^{\parallel}\ $and $g_{e}^{\perp}$ are the electron $g$-factor tensors.
The term $\mathbf{\Omega}_{imp}\cdot\mathbf{S}$ models the interaction between
the NV center electron spins and the surface impurity electron spins, via
\begin{equation}
\mathbf{\Omega}_{imp}=%
{\displaystyle\sum_{n=1}^{N}}
A_{n}\left[  S_{n,z}^{\left(  imp\right)  }\mathbf{\hat{z}}+\frac{1}{\lambda
}\left(  S_{n,x}^{\left(  imp\right)  }\mathbf{\hat{x}}+S_{n,y}^{\left(
imp\right)  }\mathbf{\hat{y}}\right)  \right]  .
\end{equation}
Here $S_{n}^{\left(  imp\right)  }$ is the spin operator of the $n-$th surface
impurity electron, $A_{n}$ is the coupling constant between the NV center
electron and the $n-$th surface impurity electron, and $\lambda$ is the
anisotropy parameter. The total number of surface impurity electrons is $N$.
We take the $z$-axis to be the normal vector to the surface plane and place
the origin at the NV center. We write the overall effective frequency as a
linear superposition of $\mathbf{\Omega}_{B}$\ and $\mathbf{\Omega}_{imp}%
$\ which we define as $\mathbf{\Omega}\equiv\left(  \mathbf{\Omega}%
_{B}+\mathbf{\Omega}_{imp}\right)  $.

We take a semiclassical approach by using the Langevin method applied to the
Bloch equation \cite{GI2012, LAX66}, we describe the stochastic dynamics of
the spin fluctuation $S\left(  t\right)  $ as
\begin{equation}
\frac{\partial\mathbf{S}\left(  t\right)  }{\partial t}=\mathbf{\Omega}%
\times\mathbf{S}\left(  t\right)  -\sum_{j=1}^{N}\left[  W_{oj}\mathbf{S}%
\left(  t\right)  -W_{jo}\mathbf{S}\left(  t\right)  \right]  -\nu
_{s}\mathbf{S}\left(  t\right)  . \label{Bloch}%
\end{equation}
Here $\nu_{s}$ is the impurity electron spin relaxation frequency. It is the
reciprocal of $\tau_{s}$, the relaxation time due to the environment,
excluding the NV center spin--impurity spin interaction. The quantities
$W_{jo}$ are the impurity hopping rates from the site $j$ to site $o$,\ where
$o$ is located right above the NV center.

\subsection{Correlation and noise spectrum\label{subsec_surfaceSpectrum}}

The noise spectrum may be written in terms of the two-time correlation
function as%
\begin{equation}
S\left(  \omega\right)  =%
{\displaystyle\int}
\left\langle S\left(  t+\tau\right)  S\left(  t\right)  \right\rangle
e^{i\omega\tau}d\tau. \label{S_noise_corr}%
\end{equation}
Defining $W_{0}=1/\tau_{c}$ where $\tau_{c}$ is the impurity electron
correlation time for the given surface terminating atom and using
Eqs.\ (\ref{Bloch}) and (\ref{S_noise_corr}) together with spin fluctuation
theory \cite{GI2012, LAX66, SMI2014, LL1981}, we obtain
\begin{equation}
\left(  -i\omega+\nu_{s}+W_{0}\right)  \mathbf{S}+\mathbf{S}\times
\mathbf{\Omega}=\frac{1}{4}+\frac{W_{0}}{N}\mathbf{S}\left(  t\right)  .
\label{FluctTheory}%
\end{equation}
The cross product in Eq.\ (\ref{FluctTheory}) can be represented using
$\mathbf{S}\times\mathbf{\Omega=}%
{\displaystyle\sum\limits_{j,k}}
\epsilon_{ijk}\Omega_{ij}S_{ij}$ where $\epsilon_{ijk}$ is the Levi-Civita
tensor and $i,j,k=x,y,z$. Then, we obtain the noise spectrum%
\begin{equation}
S\left(  \omega\right)  =\frac{\tau_{\omega}}{4}\frac{\mathcal{A}\left(
\tau_{\omega}\right)  }{1-W_{0}\tau_{\omega}\mathcal{A}\left(  \tau_{\omega
}\right)  }+c.c.,
\end{equation}%
\begin{equation}
\tau_{\omega}=\frac{1}{\nu_{s}+W_{0}-i\omega},
\end{equation}
and%
\begin{equation}
\mathcal{A}\left(  \tau_{\omega}\right)  =\frac{1}{N}\sum_{i}^{N}%
\frac{1+\Omega_{iz}^{2}\tau_{\omega}^{2}}{1+\Omega_{i}^{2}\tau_{\omega}^{2}}.
\end{equation}
In general the spin-spin interaction will be anisotropic because of the
dipolar interaction. However, the anisotropic part is averaged out in the
molecular motion in solution, and so we can focus on the fluctuations in the
$z$-direction. The average contributions of the impurities to the spin
precession, $\Omega_{i},$ in each direction along the surface will obey a
Gaussian distribution, $\mathcal{F}\left(  \Omega_{imp}\right)  $, giving us
\cite{GI2012}%
\begin{equation}
\mathcal{F}\left(  \Omega_{imp}\right)  =\frac{\lambda}{\left(  \sqrt{\pi
}\delta_{e}\right)  ^{3}}\exp\left(  -\frac{\Omega_{imp}^{2}}{\delta_{e}^{2}%
}\right)  ,
\end{equation}
where $\delta_{e}$ is the Larmor frequency of the NV center electron affected
by the impurity electron spin field and for electrons $\lambda=1$. Then
$\mathcal{A}\left(  \tau_{\omega}\right)  $ becomes%
\begin{equation}
\mathcal{A}\left(  \tau_{\omega}\right)  =\frac{1}{3}+\frac{4}{3\left(
\delta_{e}\tau_{\omega}\right)  ^{2}}-\frac{4\sqrt{\pi}\exp\left[  \frac
{1}{\delta_{e}^{2}\tau_{\omega}^{2}}\right]  }{3\left(  \delta_{e}\tau
_{\omega}\right)  ^{3}}erfc\left(  \frac{1}{\delta_{e}\tau_{\omega}}\right)  .
\end{equation}

\section{Bulk impurity interactions\label{Sec_Bulk}}

\subsection{Model\label{subsec_model}}

\begin{figure}[ptb]
\begin{center}
\includegraphics[width=8 cm] {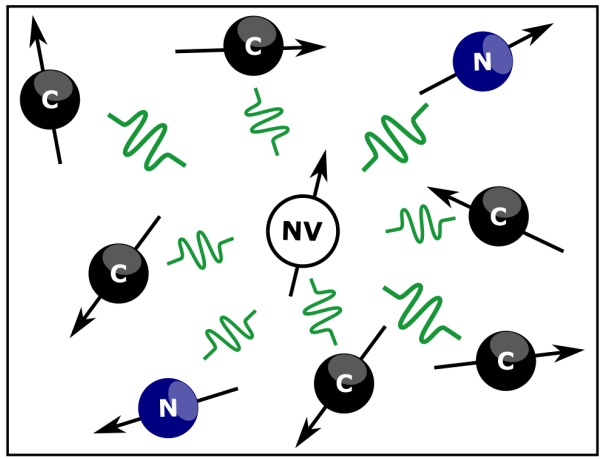}
\end{center}
\caption{The model of interacting NV center and impurities. Carbon-$13$ is
shown black, nitrogen in blue, and the dipole-dipole interaction between the
impurities and the NV center in green. The spins of the impurities are
randomly oriented.}%
\label{fig2_bulkmodel}%
\end{figure}

We now turn to the magnetic field noise due to bulk impurity interactions. The
main bulk impurities responsible for magnetic field noise are naturally
occurring carbon-$13$ isotopes and nitrogen atoms that are left over from the
process of creating the NV centers. Magnetic field noise arises from the
dipole-dipole interaction between the nuclear spins of the bulk impurities and
the electron spin of the NV center diamond. The free electron in the NV center
feels the magnetic field generated by the randomly oriented nuclear spins of
the bulk impurities. The nuclear spin magnetic field fluctuates causing the
electron to precess. The model is shown in Fig.\ \ref{fig2_bulkmodel}.

The Hamiltonian of this model is
\begin{equation}
H=H_{0}+H_{\mathrm{II}}+H_{\mathrm{SI}}. \label{Hamiltonian}%
\end{equation}
Here $H_{0}$ includes the free Hamiltonian, the zero field splitting term,
spin orbit (SO) coupling term, and the Zeeman term. It has been reported that
zero field splitting decouples the interaction between the NV center electron
spin and the spin bath and maintains coherence in NV center applications
\cite{Sekiguchi16}, thus, the noise contribution from the zero field splitting
is not significant, and we have incorporated the term into $H_{0}$.
$H_{\mathrm{II}}$ reflects the magnetic dipole-dipole hyperfine interactions
among the bulk impurities; and $H_{\mathrm{SI}}$ accounts for the
dipole-dipole hyperfine interactions between the NV center and the bulk
impurities. The two interaction Hamiltonians $H_{\mathrm{II}}$ and
$H_{\mathrm{SI}}$ are mainly responsible for generating magnetic noise. The
Hamiltonian for the magnetic dipole-dipole interactions among the bulk
impurities is%
\begin{equation}
H_{\mathrm{II}}=\sum_{j<i}\frac{b}{r_{ij}^{3}}\left[  \vec{I}_{i}\cdot\vec
{I}_{j}-3\frac{(\vec{I}_{i}\cdot\vec{r}_{ij})(\vec{r}_{ij}\cdot\vec{I}_{j}%
)}{r_{ij}^{2}}\right]  . \label{HII}%
\end{equation}
Here%
\begin{equation}
b=\frac{\mu_{0}\mu_{\mathrm{I}}^{2}}{4\pi\hbar}; \label{eq:b}%
\end{equation}
$I_{i}$ and $I_{j}$ are the nuclear spin operators for the $i$th and $j$th
nuclear spins in the cluster, $\mu_{0}$ is the vacuum permeability,
$\mu_{\mathrm{I}}$ is the nuclear magnetic moment, and $r_{ij}$ is the
distance between the nuclear spins.

The Hamiltonian for the dipole-dipole interaction between the NV center
electron spin, $\vec{S}$ and the bulk impurity nuclear spins, $\vec{I}_{i}$
with $i=1\ldots N$, is described by%
\begin{equation}
H_{\mathrm{SI}}=\sum_{i}^{N}\frac{a}{R_{i}^{3}}\left[  \vec{S}\cdot\vec{I}%
_{i}-3\frac{(\vec{S}\cdot\vec{R}_{i})(\vec{R}_{i}\cdot\vec{I}_{i})}{R_{i}^{2}%
}\right]  ,\label{HSI}%
\end{equation}
where%
\begin{equation}
a=\frac{\mu_{0}\mu_{\mathrm{e}}\mu_{\mathrm{I}}}{4\pi\hbar},\label{eq:a}%
\end{equation}
$\mu_{\mathrm{e}}$\ is the electron magnetic moment, and $R_{i}$ is the
distance between the NV center and the impurity spin.

\subsection{Correlation and noise spectrum\label{subsec_bulk2timecorr}}

\subsubsection{Two-time correlation for one
cluster\label{subsubsec_onecluster}}

We follow Hall, \textit{et al}.~\cite{hall2014analytic} and use the
correlated-cluster expansion method to calculate the two-time correlation of
the magnetic field fluctuations. We set the $z-$direction to be the crystal
axis. The magnetic field operator of the impurity cluster felt by the NV
center can be written in terms of the axial components of the nuclear
dipole-dipole hyperfine interactions as
\begin{equation}
\widehat{B}={\sum_{n=1}^{k}}\left(  T_{zx}^{n}I_{x}^{n}+T_{zy}^{n}I_{y}%
^{n}+T_{zz}^{n}I_{z}^{n}\right)  , \label{Bsum}%
\end{equation}
where $T_{zj}$ with $j=x,y,z$\ is the hyperfine interaction tensor, $I$\ is
the impurity nuclear spin, and $k$ is the number of spins within the cluster.
When the applied magnetic field is large, the secular (S) approximation which
is similar to the rotating-wave approximation allows us to neglect the non
energy conserving off-diagonal terms of the nuclear dipole-dipole Hamiltonian
(Eq.\ (\ref{HII})). For small applied magnetic fields, the non-secular method
(NS), all possible terms of Eq.\ (\ref{HII}) must be considered. For the
purposes of this study we consider anything below $1$ \textrm{G} to be low
fields and anything above $100$ \textrm{G} to be large fields.

Starting from Eq.\ (\ref{Bsum}) we calculate the two-time correlation function
of the magnetic field of the impurity cluster for both the secular and
non-secular approximation. Here we consider a cluster of $k=2$ spins because
Hall \textit{et al}. \cite{hall2014analytic} shows that clusters of $k>2$
spins exhibit identical properties as the $k=2$ spin cluster due to rapid
falloff of dipole-dipole coupling strength. In the secular approximation, the
magnetic field two-time correlation along the crystal axis can then be written
as%
\begin{equation}
\left\langle B(t)B(0)\right\rangle _{\mathrm{S}}=T_{z,1}^{2}+T_{z,2}%
^{2}-\left[  T_{z,1}-T_{z,2}\right]  ^{2}\sin^{2}\left(  \frac{B_{12}t}%
{2}\right)  ; \label{Correl-Sfinal}%
\end{equation}
in the non-secular calculation, the autocorrelation function of the axial
magnetic field will be%
\begin{equation}
\left\langle B(t)B(0)\right\rangle _{\mathrm{NS}}=\left(  T_{z,1}^{2}%
+T_{z,2}^{2}\right)  \left[  1-\frac{3}{4}\sin^{2}\left(  \frac{3B_{12}t}%
{4}\right)  \right]  . \label{Correl-NSfinal}%
\end{equation}
Here $B_{12}=b/r_{12}^{3}\left[  1-3\cos^{2}\left(  \theta_{12}\right)
\right]  $ is the axial flipping rate of the cluster magnetic field between
being parallel or anti-parallel to the $z-$axis. $B_{12}$ is determined by
$r_{12}$ the distance between the two spins of the cluster, and $\theta_{12}$
the angle between $r_{12}$ and the $z-$axis.

\subsubsection{Two-time correlation function for $N$
clusters\label{subsubsec_Nclusters}}

To generalize the result of Sec.\ \ref{subsubsec_onecluster} to the $N$
cluster case, we must sum over all the clusters in the system. Since the
dynamics of the clusters are stochastic in nature and their magnetic field
strengths will depend on their radial distribution, we use a probability
density function of finding a fixed number $n^{\prime}$of spins on a
concentric sphere within a distance $r^{\prime}$of the NV center. The
probability density function for considering a single cluster will follow a
Poisson distribution and will be as follows,%

\begin{equation}
p(r^{\prime})=\left(  4\pi n^{\prime}r^{\prime2}\right)  \exp\left[
-\frac{4\pi n^{\prime}r^{\prime3}}{3}\right]  . \label{p(r')}%
\end{equation}
Now if we consider the joint probability distribution for $N$ clusters, the
probability density function for $N$ clusters with $k$ spins will become,%
\begin{equation}
P(r^{\prime},\ldots,r_{k}^{^{\prime}})=%
{\displaystyle\prod\limits_{j=1}^{k}}
p_{r}(r_{j}^{\prime}), \label{p(r'_to_r'k}%
\end{equation}
where $p_{r}(r_{j}^{\prime})$ is the probability density function of a cluster
at a distance $r_{j}^{\prime}$ where $j=1\ldots k$. With Eqs.\ (\ref{p(r')})
and (\ref{p(r'_to_r'k}), we obtain
\begin{equation}
P(r^{\prime})=\frac{\left(  4\pi n^{\prime}r_{k}^{\prime2}\right)  }%
{(k-1)!}\left(  \frac{4\pi n^{\prime}r_{k}^{\prime3}}{3}\right)  ^{k-1}%
\exp\left[  -\frac{4\pi n^{\prime}r_{k}^{\prime3}}{3}\right]  .
\end{equation}
Then the two-time correlation function of the system is as follows%
\begin{equation}
C(t)=%
{\displaystyle\sum\limits_{i=1}^{N}}
\langle B(t)B(0)\rangle_{i}P(r^{^{\prime}}), \label{2t-corr}%
\end{equation}
and the noise spectrum is
\begin{equation}
S(\omega)=\int_{0}^{t}C(t^{\prime})\cos(\omega t^{\prime})dt^{\prime}.
\label{BNoiseSpec}%
\end{equation}
Plugging in Eqs.\ (\ref{Correl-Sfinal}) and (\ref{Correl-NSfinal}) to
Eq.\ (\ref{2t-corr}), we can work out our magnetic field correlation functions
for $N$ clusters and have the following form%
\begin{equation}
C_{\mathrm{S}}(t)=\frac{8}{5}\left(  \frac{4}{3}\pi an\right)  ^{2}[1-\frac
{1}{3}M(t)], \label{CsFunct}%
\end{equation}
for the secular approximation and%
\begin{equation}
C_{\mathrm{NS}}(t)=\left(  \frac{8}{3}\pi an\right)  ^{2}[1-N(t)],
\label{CnsFunct}%
\end{equation}
for the non-secular case. Here $n$ is the number density of the impurities,
and $M(t)$ and $N(t)$ are the cluster magnetization functions of the system
and are given by \cite{hall2014analytic}
\begin{equation}
M(t)=\frac{4\pi\sqrt[3]{6}}{\Gamma(\frac{8}{3})}(\pi bnt)^{5/3}-\frac{8\pi
}{\sqrt{3}\Gamma(\frac{4}{3})}(\pi bnt)^{2}, \label{M(t)}%
\end{equation}
and%
\begin{equation}
N(t)=\left(  \frac{4}{9}\pi^{2}bnt\right)  , \label{N(t)}%
\end{equation}
where $\Gamma$ is the gamma function. Plugging Eqs. (\ref{CsFunct}) and
(\ref{CnsFunct}) into Eq. (\ref{BNoiseSpec}), the noise spectrum becomes,
\begin{equation}
S(\omega)_{\mathrm{S}}=\int_{0}^{t}\frac{8}{5}\left(  \frac{4}{3}\pi
an\right)  ^{2}[1-\frac{1}{3}M(t^{\prime})]\cos(\omega t^{\prime})dt^{\prime},
\label{noise_secular}%
\end{equation}
in the secular approximation and%
\begin{equation}
S(\omega)_{\mathrm{NS}}=\int_{0}^{t}\left(  \frac{8}{3}\pi an\right)
^{2}[1-N(t^{\prime})]~\cos(\omega t^{\prime})dt^{\prime},
\label{noise_nonsecular}%
\end{equation}
in the non-secular approximation.

\section{Results and discussions\label{Sec_results}}

\subsection{Surface impurity noise\label{subsec_SurfaceNoiseDiscussion}}

\begin{figure}[ptb]
\centering
\includegraphics[width=7 cm]{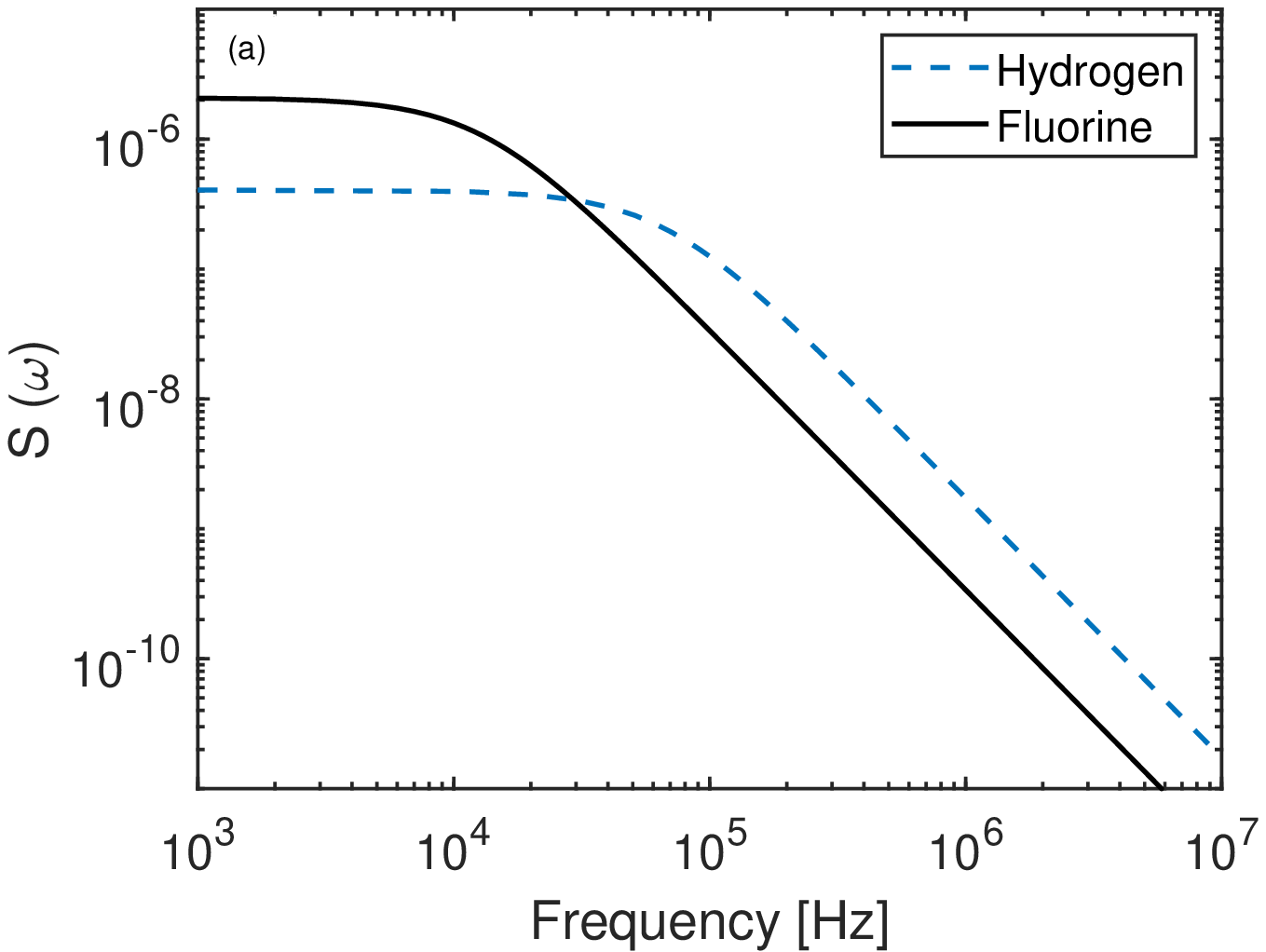}
\includegraphics[width=7 cm]{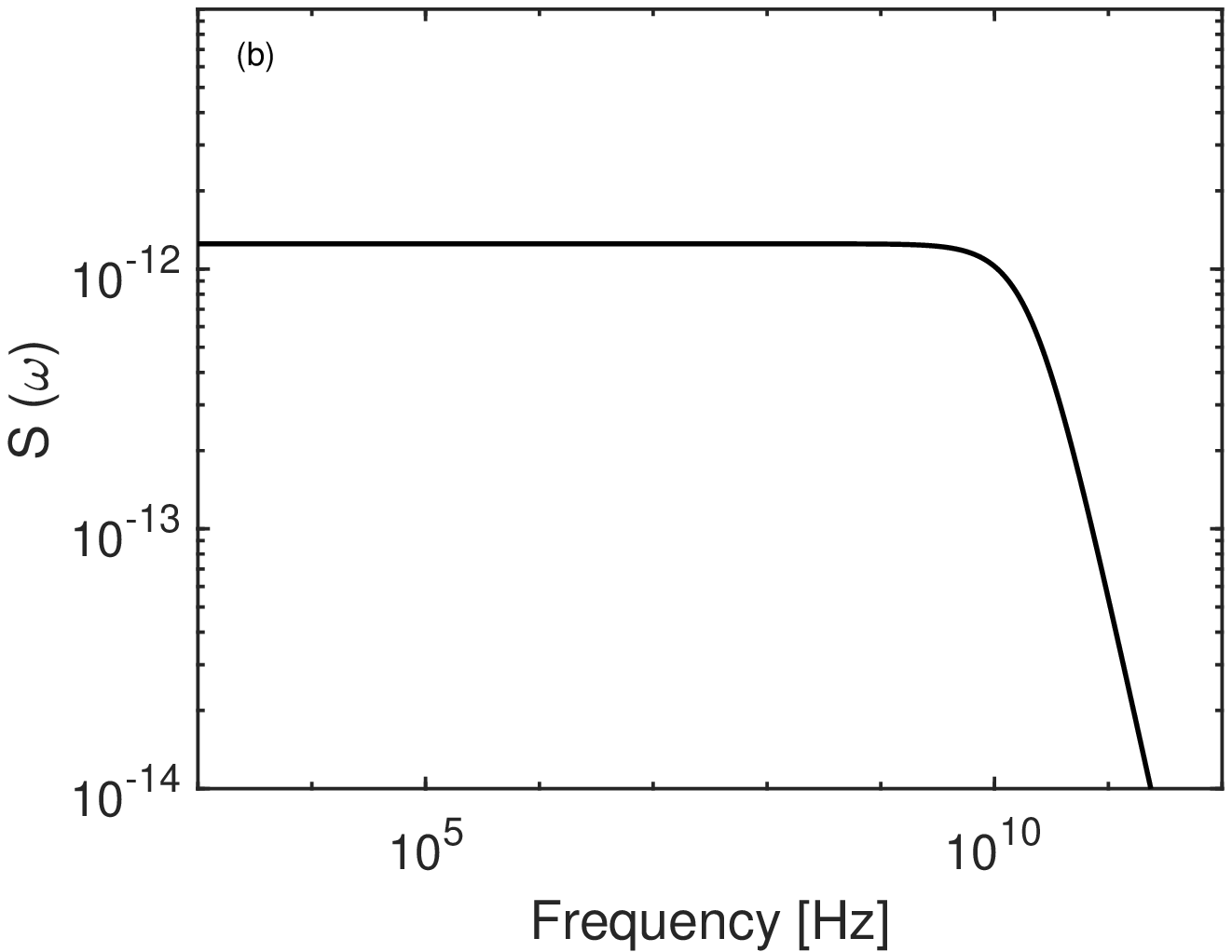}\caption{Noise spectrum of
(a) F- and H-terminated surfaces and (b) O-terminated surface with $N=10^{6}$
paramagnetic surface impurities. The O-terminated system has much less noise
than the H- and F- terminated systems. This is because oxygen electron spins
have an extremely short spin precession relaxation time.}%
\label{fig:surfnoise}%
\end{figure}

Figure \ref{fig:surfnoise} (a) and (b) show the calculated noise spectral
density for H-, O-, and F- terminated NV center diamonds. The constant noise
floor at low frequency comes from the electron spins of the surface impurities
and the NV center randomly precessing and trying to relax to an equilibrium
state. At high frequencies, the noise spectrum follows a power law.

\subsubsection{Spin flip noise and spin precession noise}

At low frequencies, F-terminated surface generates the highest noise of the
three termination systems. At around $40$ \textrm{kHz} and above, the
H-terminated surface has higher noise than the F-terminated surface. The
O-terminated surface has much less noise than H- and F-terminated surfaces in
the practical operational frequency range ($10^{3}-10^{8}$ \textrm{Hz}). At
low frequencies ($<10^{4}$ \textrm{Hz}), the difference is as much as $5$ or
$6$ orders of magnitude.

This huge difference in magnetic field noise can be attributed to the very
short ($\sim7.5$ $\mathrm{ps}$) spin-lattice relaxation time of the impurity
electron spin in the O-terminated surface compared to that of H- and
F-terminated surface, where the spin-lattice relaxation times are on the order
of microseconds: $13$ $\mathrm{\mu s}$ for hydrogen and $300$ $\mathrm{\mu s}$
for fluorine. A short relaxation time ($\delta_{e}<\nu_{s}$)\ leads to a short
impurity-impurity correlation time, and thus, the noise is purely due to spin
flip fluctuations since the much quicker spin-precession will not contribute
to noise while they are correlated. On the other hand, long relaxation time
($\delta_{e}>\nu_{s}$) leads to a long impurity-impurity spin correlation
time, which in turn adds the effect of spin-precession fluctuation noise to
the spin flip fluctuation noise. The huge differences in the amount of noise
by termination suggest that spin precession fluctuations seem to dominate the
magnetic field noise.

\subsubsection{Impurity number and noise}

The effective relaxation time is limited by the faster of the spin flips and
the hopping rates. The number of impurities affects only the hopping rate.
Thus if $\nu_{s}<$$W_{0}$, the effective relaxation time is limited by the
hopping rate, while if $\nu_{s}>W_{0}$, the relaxation time is limited by the
spin flip process. The hopping induced spin flip rate depends on the number of
impurity spins. For O-terminated surfaces, $\nu_{s}\gg W_{0}$, and so the
noise is not affected by the number of impurities. However, H-\ and
F-terminated surfaces are different. At low $N$, the relaxation times of H-
and F-terminated surfaces are determined by the hopping rate, since $\nu_{s}<
$$W_{0}$ in this regime. At high $N$, the situation is reversed: the spin flip
rate dominates the relaxation times, since $\nu_{s}>W_{0}$. For H-terminated
surfaces, the transition between the low $N$ and high $N$ regimes occurs
around $N=10^{6}$.

\begin{figure}[ptb]
\centering
\includegraphics[width=0.5\linewidth]{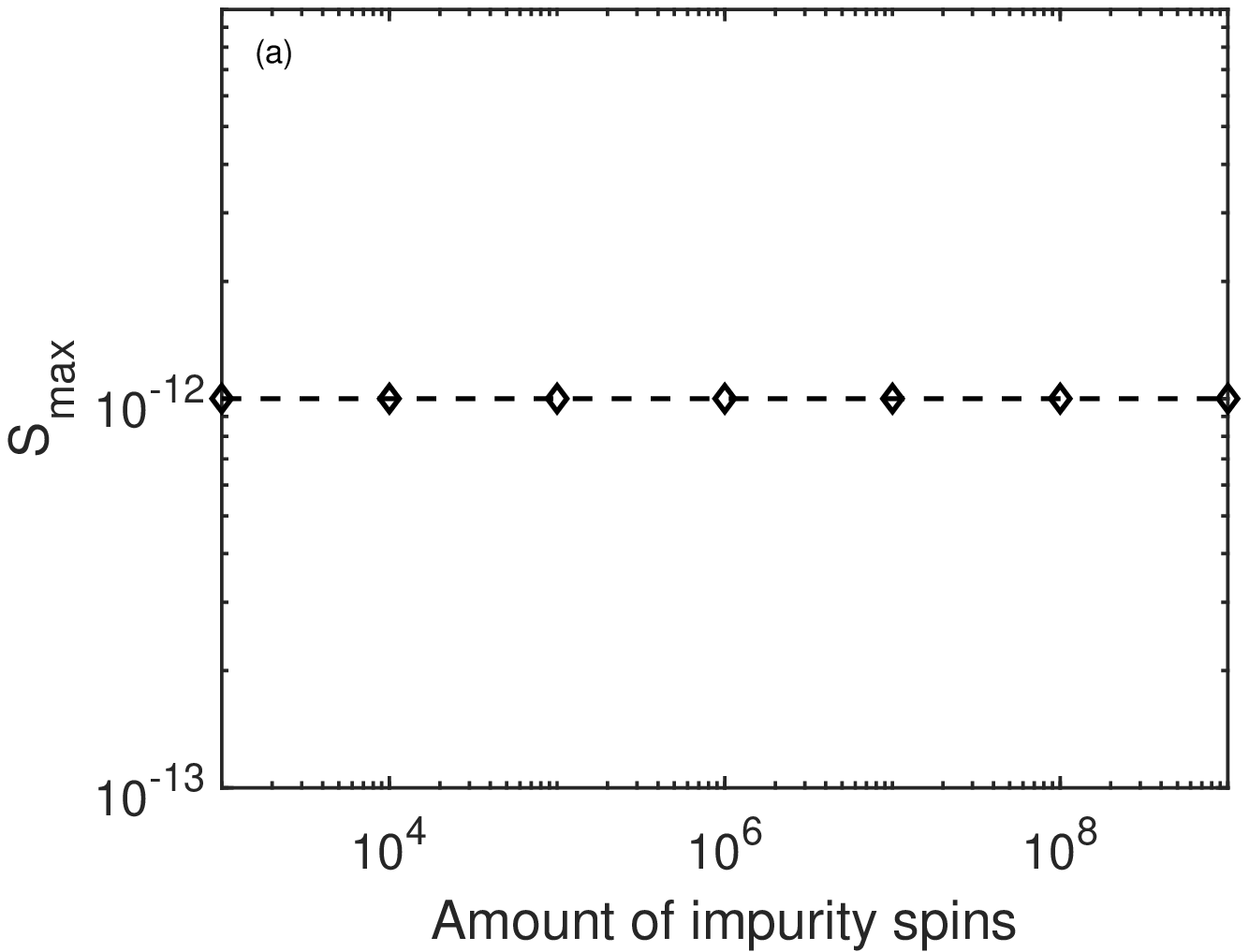}
\includegraphics[width=0.5\linewidth]{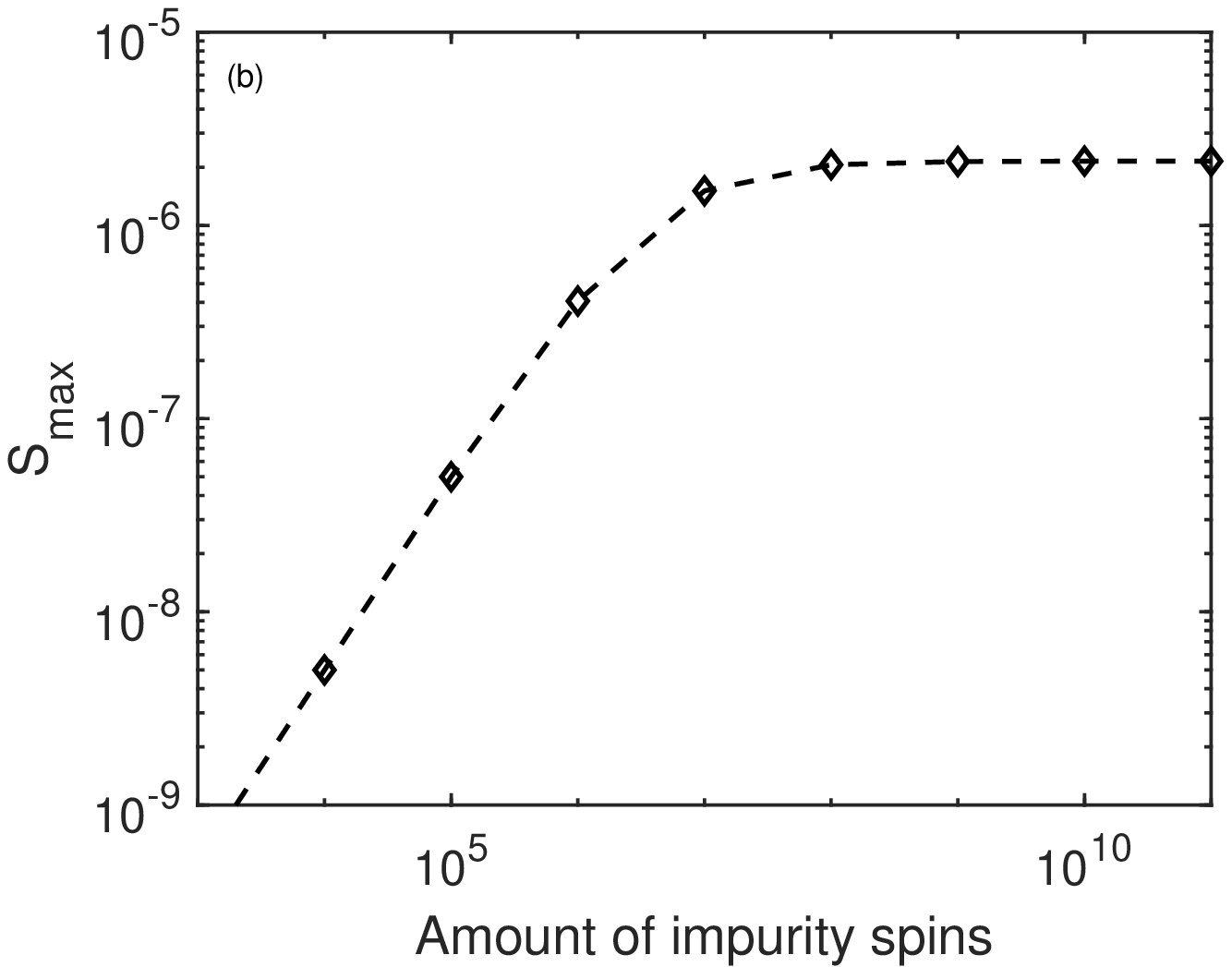} \caption{Noise at
$\omega=10^{3}$ \textrm{Hz}\ (the largest value of the noise spectrum) vs. the
number of impurities for (a) O-terminated surface and (b) H-terminated
surface. Calculated values are shown in diamonds and dotted lines are
interpolation. The the noise spectrum is unaffected by the number of
impurities in the O-terminated surface, while in the H-terminated surface, the
amount of noise increases with the number of impurities until reaching a
relaxation time controlling mechanism transition occurs at about $N=10^{6}$.}%
\label{fig5_concentration}%
\end{figure}

Figure \ref{fig5_concentration} shows the noise at $\omega=10^{3}$ Hz (i.e.,
the largest noise value since the low frequency region has almost flat region
in the noise spectrum as seen in Fig.\ \ref{fig5_concentration}) vs. the
number of impurity spins for H-\ and O-terminated surfaces. In O-terminated
systems, where the relaxation time is limited by spin flips, the noise
spectrum exhibits no impurity number dependence. In H-terminated systems,
however, we see the transition from the low-$N$ regime in which relaxation
time is controlled by hopping ($\nu_{s}<W_{0}$) to the high-$N$ regime in
which spin flips control the relaxation time ($\nu_{s}>W_{0}$). This
transition indeed occurs around $N\sim10^{6}$ as the relation between $\nu
_{s}$ and $W_{0}$ shifts to $\nu_{s}>W_{0}$.

\subsection{Bulk impurity noise spectrum\label{subsec_BulkNoiseDiscussion}}

\begin{figure}[ptb]
\centering
\includegraphics[width=7 cm]{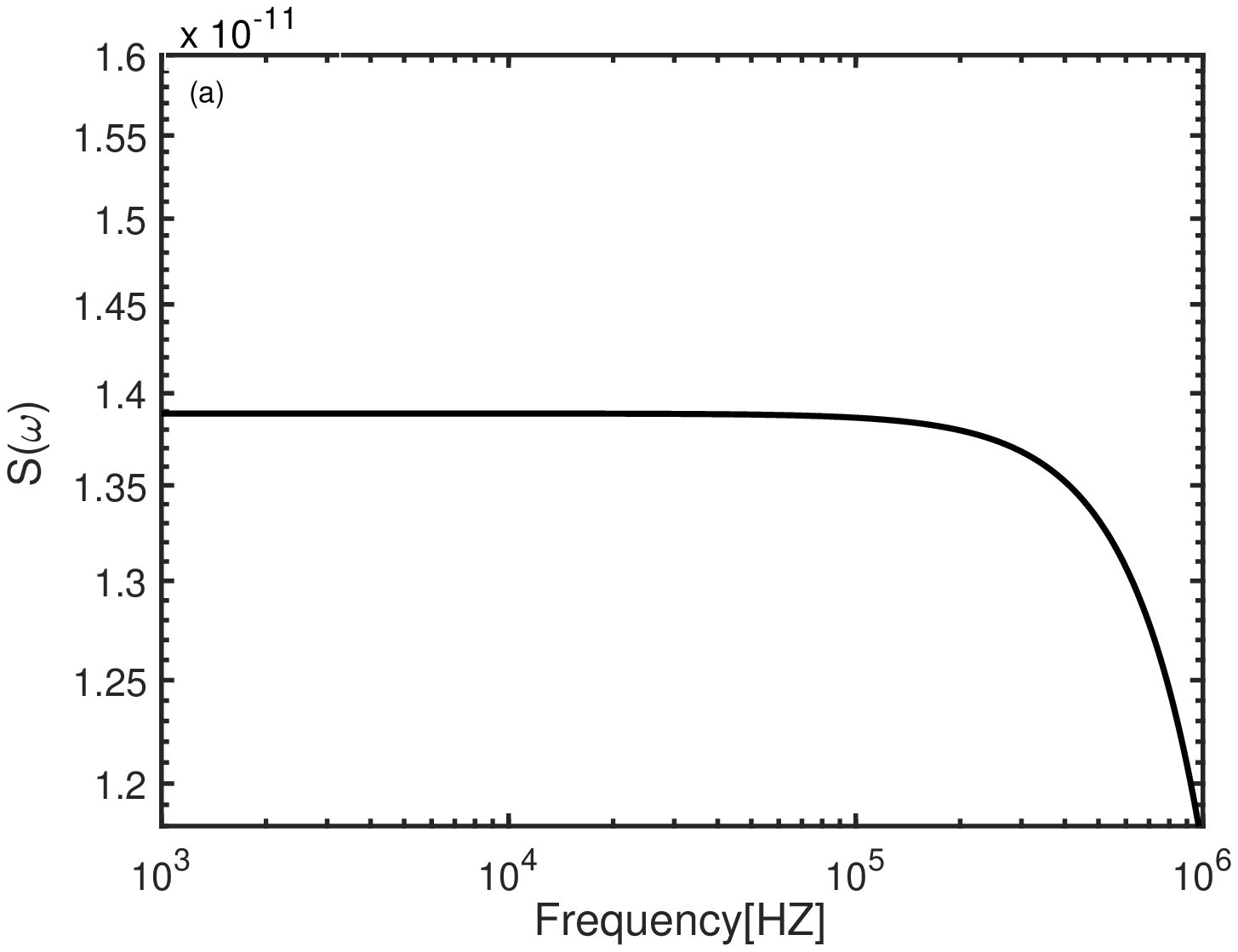}
\includegraphics[width=7 cm]{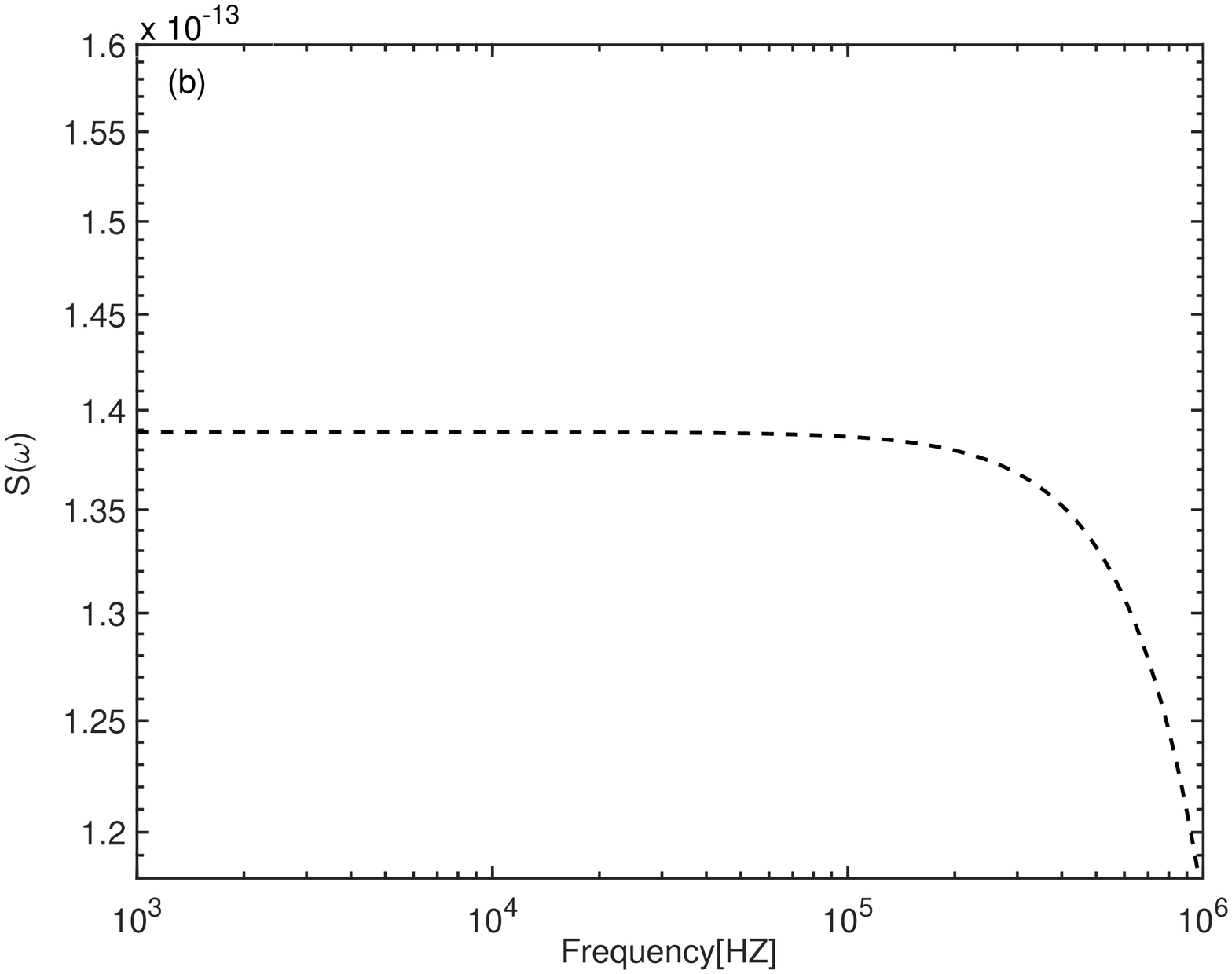}\caption{magnetic noise spectrum
for (a) carbon-$13$ impurities and (b) nitrogen impurities. The calculation is
done at low applied magnetic field, using the non-secular method. The assumed
number density is $10^{19}\mathrm{atoms}/\mathrm{cm}^{3}$ for carbon-$13$ and
roughly $10^{18}\mathrm{atoms}$/$\mathrm{cm}^{3}$ for nitrogen. These are
typical experimental number densities. The carbon-$13$ noise is roughly $100$
times greater than that of nitrogen.}%
\label{fig6_C_N_NonSec}%
\end{figure}

Figure \ref{fig6_C_N_NonSec} shows the noise spectrum produced by (a)
carbon-$13$ and (b)nitrogen impurities, calculated in the non-secular method.
The results show that the noise due to dipole-dipole interactions between the
NV center and carbon-$13$ impurities is roughly two orders of magnitude
greater than that the noise from the nitrogen impurities. The reason for the
difference is number density, which affects the noise spectrum in two ways.
(1) Because the number density of carbon-$13$ is greater than that of
nitrogen, the average interaction distance between carbon-$13$ atoms and the
NV centers is smaller than the average nitrogen-NV center interaction
distance. Consequently, the carbon-$13$ nuclear spins have stronger
interactions with the NV centers, generating more noise. (2) A greater number
density of impurities increases the number of nuclear spin flip-flops. These
spin flip-flops perturb the free electron in the NV center, causing the
electron to precess. The precession, in turn, leads to greater fluctuations of
the NV center electron spin state. The maximum noise of $S(\omega)$\ is
related to the number density by a power-law of $S(\omega)\propto n^{2}$.

\subsubsection{Secular approximation vs.~non-secular
method\label{subsec_TwoNumerical}}

\begin{figure}[ptb]
\centering
\includegraphics[width=7 cm]{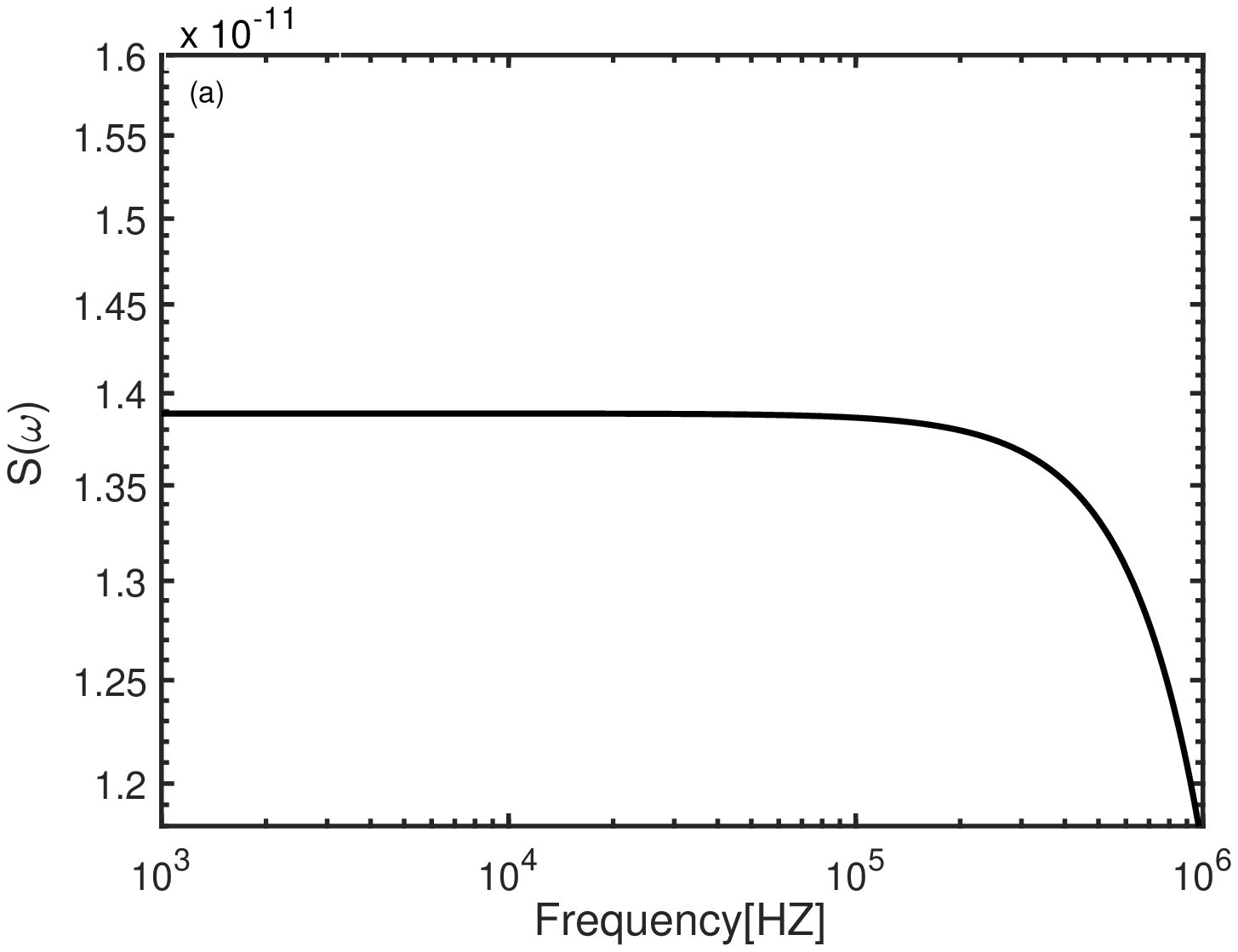}
\includegraphics[width=7 cm]{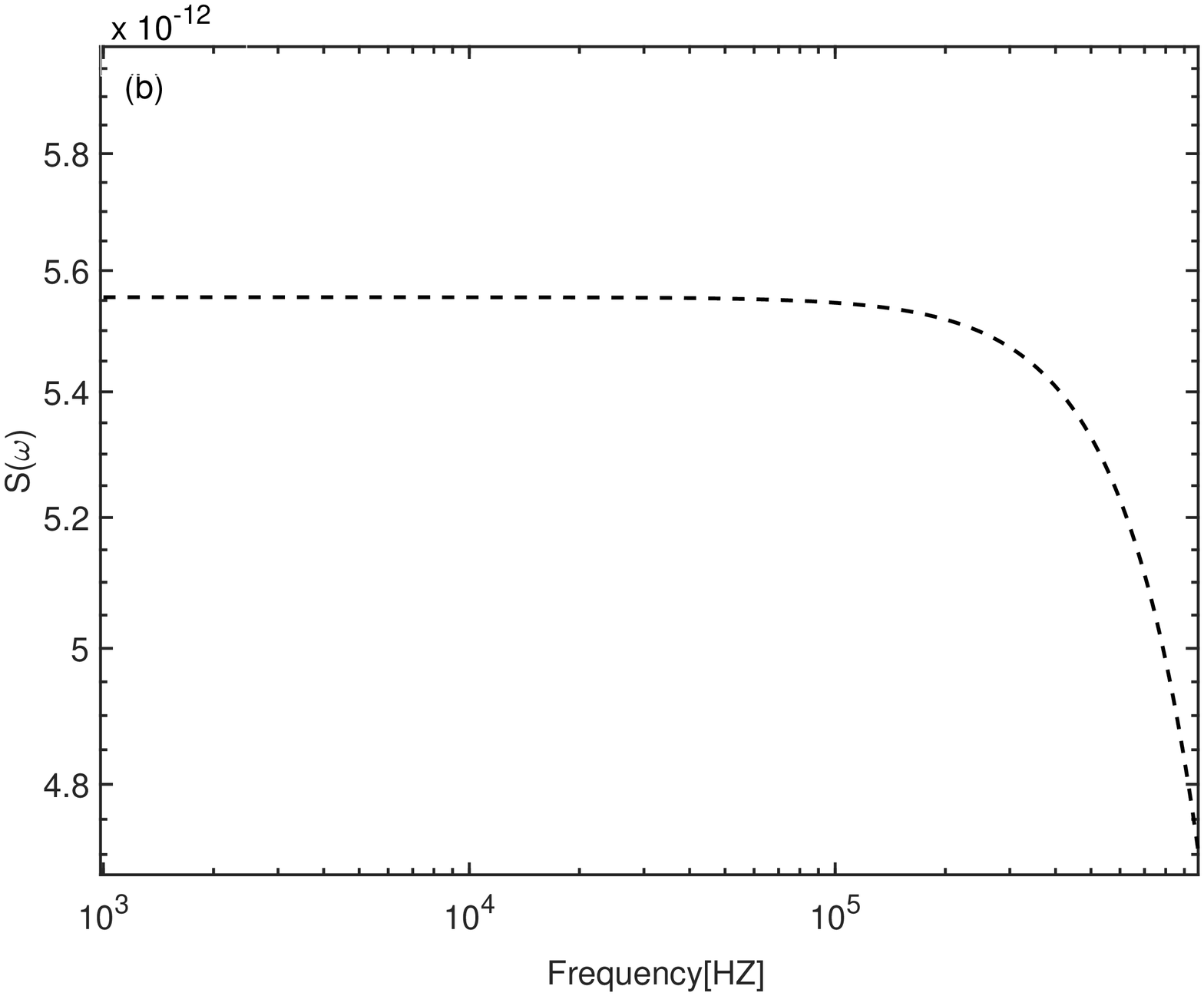}\caption{Magnetic noise spectrum
from carbon-13 impurities calculated with (a) the secular approximation and
(b) the exact non-secular method. The noise is three times larger in the
non-secular calculation. Thus the secular approximation underestimates the
noise at low applied magnetic field.}%
\label{fig7_C_S_NS}%
\end{figure}

Most calculations of bulk impurity interactions tend to use the secular
approximation, because it simplifies the analysis. However, at low applied
magnetic field, the secular approximation is inaccurate. To demonstrate this,
we calculate the noise spectrum at low magnetic field using both the secular
approximation and non-secular methods. Figure\ \ref{fig7_C_S_NS} shows the two
calculations for the noise spectrum from interactions of NV centers with
carbon-$13$. We see that the secular approximation underestimates the amount
of noise by a factor of $2.7$.

The reason for the underestimate is as follows. There is a degeneracy between
the $m_{s}=+1$ and $m_{s}=-1$\ ground states at zero applied magnetic field,
which is broken by applying a magnetic field. If the applied magnetic field is
high, the split is large enough so that the off-diagonal terms of the nuclear
dipole-dipole interaction Hamiltonian (Eq.\ (\ref{HII})) can be neglected, and
thus, the secular approximation is valid. However, at low\ applied magnetic
field, these off-diagonal elements are not negligible. In fact, they account
for our factor of $2.7$ discrepancy between the secular and non-secular
calculations. Some experimentalists deliberately work at high applied magnetic
fields so that the secular approximation will be valid for their experiments
\cite{BSS17,HKS2016}.

\subsection{Surface vs. bulk impurity noise\label{subsec_BulkvsSurface}}

Unlike surface impurities, bulk impurities are located at fixed positions,
thus there is no hopping between different sites. As a result, bulk impurity
noise is purely spin flip noise. On the other hand, the surface impurities can
hop from site to site resulting in possibly having both spin flip noise and
spin precession noise or spin flip noise only depending on either the spin
flip rate or the hopping rate controlling the effective relaxation time as
discussed in Sec.\ \ref{subsec_SurfaceNoiseDiscussion}. We have seen that an
O-terminated surface, that has only spin flip noise due to short relaxation
time, produces $5$ to $6$ orders of magnitude less noise at low frequency
range than H- and F-terminated surfaces that have both spin precession noise
and spin flip noise in the low frequency range of $<10^{4}$ \textrm{Hz}.
Comparing the noise between surface noise and bulk noise, we notice that bulk
noise is about the same or even less noise than the O-terminated, which
confirms that the spin precession noise is indeed the major magnetic noise source.

However, for the bulk impurities, the impurity number density also has a
significant effect on noise as impurities are at fixed position and the
interaction strength changes with the density while for the surface
impurities, the impurity numbers increase noise as long as the relaxation is
limited by the hopping rate as the impurity number changes the hopping rate.

\section{Conclusions\label{Sec_conclusion}}

We have\ analytically and numerically investigated the magnetic noise from
paramagnetic surface and bulk impurities. For surface impurities, we studied
the interactions between the paramagnetic surface impurity electron spins and
the NV center electron spins. In such systems, different species can terminate
the dangling bonds at the diamond surface, and the impurities are dissolved in
a condensed thin layer of water. We studied three different surface
terminations; hydrogen, oxygen, and fluorine. Applying the Langevin method to
the Bloch equation, we calculated the spin-spin noise spectrum.

Our results show that there are two mechanisms that contribute to noise: spin
flip fluctuations and spin precession fluctuations. A short relaxation time
leads to a short impurity-impurity spin correlation time resulting in a spin
flip fluctuation noise only. On the other hand, a long relaxation time leads
to a long impurity-impurity spin correlation time, which results in activating
another effect of spin-precession fluctuation noise to the already existing
spin flip fluctuation noise. Thus, an O-terminated surface, for which the
spin-lattice relaxation time is very short, have only the spin flip
fluctuation noise while H- and F-terminated surfaces, with a longer
spin-lattice relaxation time, have both spin flip and precession fluctuation
noise. The spin precession fluctuation noise is much larger than the spin flip
fluctuation noise, dominating the magnetic field noise in NV center diamonds.

The effective relaxation time of impurities is limited by the faster of either
the spin flip rate or the hopping rate. The reaction limited nature of the
bonding between the hopping terminating atoms and the dangling bond allow most
of the sites above and near the NV center to be occupied. At this point, the
terminating atoms with long relaxation times are still generating spin
precession noise which generates more noise than spin flips due to hopping. At
high frequencies (short time scales), the dominating presense of the spin
precession is not fully felt as the terminating atoms have not had the time to
make the energetically stable bond with the dangling bond. The number of
impurities affects only the hopping rate. Since the relaxation time of an
O-terminated surface is determined by the spin flip rate, it is not affected
by the change in the number of impurities. On the other hand, the relaxation
times of H-and F-terminated surfaces are controlled by either the hopping
rate, if $N<10^{6}$, or the spin flip rate, if $N>10^{6}$ and the transition
between two control mechanism can be seen clearly by plotting the maximum
noise vs impurity.

The bulk magnetic noise is caused by the bulk impurity nuclear spin
interactions with the NV center electron spin. The major bulk impurities of
the system are carbon-$13$ and the leftover nitrogen from implantation. We
used the correlated-cluster expansion to calculate the magnetic field noise
from the nuclear spin flip fluctuations. We numerically evaluated the noise
spectrum for both the carbon-$13$ and the nitrogen impurities at low magnetic
fields. We find that the noise from carbon-$13$ was roughly two orders of
magnitude larger than that from nitrogen. We attribute this difference to
number density. The number density affects the noise spectrum through two
physical mechanisms: the distance between the impurities and the NV center,
which affects the interaction strength, and the spin flip rate. Carbon-$13$
has greater number density than nitrogen, and hence produces greater noise.
Different grade of NV-center environments are used in experiment from
$99.999\mathrm{\%}$ carbon-$12$\ to $>50\mathrm{\%}$ enriched carbon-$13$.
Plotting the maximum noise versus number density, a linear increase as the
number density increases is clearly seen.

We also examined the applicability of the secular approximation that is
commonly used to estimate the noise. We find that the secular approximation
underestimates the amount of noise when the applied magnetic field is low.
This is related to lifting the degeneracy of $m_{s}=\pm1$ states. At high
applied magnetic field, the split is large enough that some of the
off-diagonal terms of the nuclear dipole-dipole interaction matrix can be
neglected, and thus the secular approximation is valid. However, at low
magnetic field, the off-diagonal elements are not negligible, and must be
considered in the noise calculations.

Comparing the surface and bulk impurity noise spectra, surface impurities with
O-terminated surface and bulk impurities have significantly less noise than
the H- and F-terminated surfaces due to only having spin flip noise and no
spin precession noise. It might be possible to employ research techniques
similar to the spin noise spectroscopy done for gallium-arsenide to
distinguish spin precession noise from spin flip noise \cite{Ryzhov16}, which
could help indicate terminating atoms with the desirable rapid spin precession
relaxation. However, it is beyond the scope of this paper.

The inclusion of the spin precession in our model has given us insight into a
mechanism other models have looked over. For the bulk impurities, the impurity
number density also has a significant effect by changing the interaction
strength. For the surface impurities, the impurity numbers increase noise as
long as the relaxation is limited by the hopping rate as the impurity number
changes the hopping rate.

We thank Jonathan Tannenhauser for the valuable input. This work is supported
by NSF DMR-1505641.




\end{document}